# All-dielectric silicon microring resonators with deeply sub-diffractive mode volumes

Saddam Gafsi and Judson D. Ryckman*

*Holcombe Department of Electrical and Computer Engineering, Clemson University, Clemson, SC 29634*

**Abstract:** We report an approach to achieve deeply sub-diffractive mode volumes in all-dielectric ring resonators. Specifically, we leverage simultaneous usage of nanostructured subwavelength metamaterial waveguides and excitation of standing wave rather than travelling wave whispering gallery mode (WGM) resonances. This principally enables the mode volume to be reduced by factors >10 – 100 compared to a diffraction limited silicon microring resonator. We also highlight how a combination of diffractive mode volume scaling and sub-diffractive near field intensity scaling indicates a lower limit for the sub-diffractive all-dielectric mode volume, $V'_{min}$, which scales in proportion to the mode order times the refractive index raised to the seventh power, i.e.: $V'_{min} \sim mn^{-7}$. Through both numerical modeling and experimental characterization, we systematically investigate the modal properties of metawaveguide ring resonators comprised of Si and SiO$_2$. The sub-diffractive devices are experimentally shown to support standing wave resonances with high intrinsic quality factors ~10$^4$-10$^5$. These metamaterial ring resonators present a promising WGM platform for interfacing wavelength scale optics with sub-wavelength matter.

## 1 Introduction

Whispering-gallery mode (WGM) resonators, including microring resonators (MRRs), are an important category of optical microcavities that support enhanced light-matter interactions by recirculating light along a closed path. Compared to other types of microcavities, such as photonic crystals and Fabry-Perot etalons, WGM resonators are often favored for their distinct combination of flexible and straightforward design, versatile input/output coupling arrangements, ease of integration, multimode nature, widely tunable modal properties, and free-spectral range. Taking advantage of these versatile attributes, WGMs have found wide adoption in applications ranging from sensing, to nanomanipulation, non-linear and quantum optics, modulation, multiplexing, filtering, photodetection, comb sources, signal processing and more [1].

To date a large body of work has been dedicated to studying, controlling, and/or improving temporal photon confinement and $Q$ factor of WGMs and MRRs, revealing important design and experimental factors that impact scattering, bending, absorptive, and coupling losses[2–10]. Recent process and design optimizations across each of these characteristics has led to the realization of record Q factor WGM resonators with $Q$s ~10$^6$ to 10$^9$ across a variety of integration formats and materials [5,11–14].

While numerous works have advanced the limits of $Q$ factor in WGM resonators and MRRs, comparatively fewer works have investigated methods to reduce the optical mode volume $V$ of such resonators [15]. For a high $Q$ dielectric resonator, the electric mode volume may be classically defined according to:

$$V = \frac{\int \epsilon(r)|E(r)|^2 dV}{\epsilon(r_{max})|E(r_{max})|^2} = \frac{\int U_e dV}{\max[U_e]} \quad (1)$$

where $U_e$ represents the electric field energy density. More generalized expressions for the mode volume of open optical systems are necessary for the analysis of highly dissipative resonators [16–18]. Per Eq. (1), the mode volume $V$ is minimized when the peak electric field energy density (normalized to the total electric field energy) is maximized.

Reducing $V$ is advantageous for numerous applications which demand improved optical concentration or local field enhancement. A small mode volume is particularly beneficial for interfacing wavelength scale optics with subwavelength or nanoscale objects, for example in atom-light interfaces[19], quantum light sources[20–22], efficient modulators[23–25], and nanoparticle sensors and traps[26–28]. More broadly, achieving low $V$ is a key

---

***Judson D. Ryckman:** Clemson University, Clemson, SC, USA; jryckma@clemson.edu;
**Saddam Gafsi:** Clemson University, Clemson, SC, USA; sgafsi@clemson.edu;

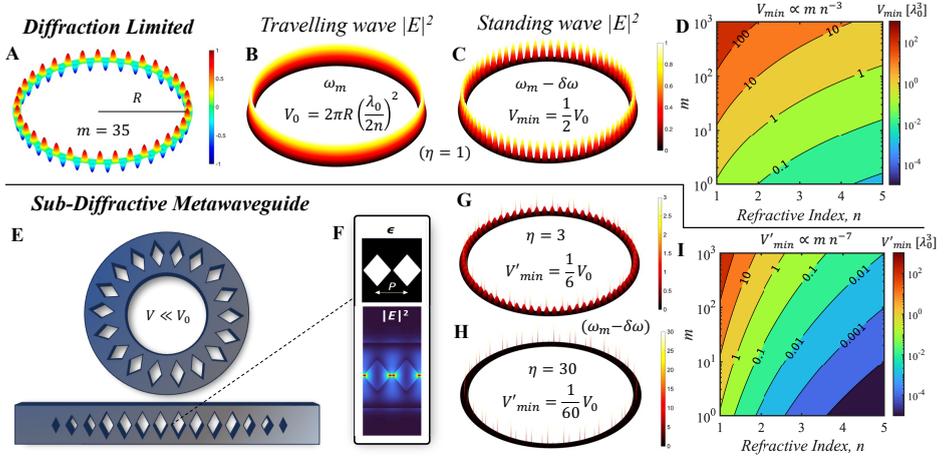

**Figure 1: The microring cavity and the resonant mode properties** (A) Diffraction limited mode profile for $m = 35$. (B) Time averaged electric field intensity profile for travelling wave and (C) standing wave resonances alongside their respective mode volumes. (D) Contour map revealing the $mn^{-3}$ mode volume scaling of diffraction limited WGMs. (E) Illustration of a metawaveguide microring resonator comprised of low index diamonds in each unit cell. (F) Visualization of the dielectric function and electric field intensity, which is strongly enhanced in the narrow dielectric bridge between adjacent diamonds. (G, H) Conceptual visualization of a standing wave WGMs with periodically distributed field intensity enhancements. (I) Contour map revealing the $mn^{-7}$ scaling of sub-diffraction limited WGMs.

factor in maximizing the overall spatiotemporal confinement, as expressed by the $Q/V$ figure of merit and Purcell factor [16,29].

Here, we introduce and experimentally demonstrate an approach for achieving sub-diffraction limited optical mode volumes $V$ in all-dielectric MRRs. Our approach avoids the use of plasmonic field enhancement[15], which often comes at the cost of elevated optical losses, and instead utilizes all-dielectric field enhancement achievable in high-index-contrast systems structured on the subwavelength scale[30]. Specifically, we employ subwavelength metamaterial waveguides with diamond unit cells, which simultaneously lifts the travelling wave degeneracy to form standing wave whispering gallery modes, and locally enhances the electric field energy density by a maximum theoretical factor approaching $(n_{hi}/n_{lo})^4$. We demonstrate oxide cladded Si MRRs with reduced mode volumes that are ~5-10x below the diffraction limit and ~10-20x smaller than travelling wave resonators of the same radius. We predict the ability to reduce the mode volume by >100x in suitably high index contrast systems. Through experimental spectral characterization, we assess the prevalence of standing, partial standing, and travelling wave modes and identify important trade-offs and parameters that influence both $Q$ and $V$. Our results provide insight into the design of ultra-low $V$ MRRs and serves as an important step toward metamaterial enabled high $Q/V$ WGM resonators -- devices that hold wide ranging applications from sensing to quantum light sources.

## 2 Approach

The optical mode volume for light resonating in a circular loop resonator formed from a diffraction limited optical waveguide with cross-sectional mode area $(\lambda_0/2n)^2$, refractive index $n$, and radius $R$, as illustrated in Fig. 1B, can be approximated as:

$$V_0 = 2\pi R \left(\frac{\lambda_0}{2n}\right)^2 = 4\pi m \left(\frac{\lambda_0}{2n}\right)^3 \qquad (2)$$

in the case of a *travelling* WGM, where $m$ is the mode order, describing the integer number of wavelengths $\lambda_0/n$ constructively circulating in a single loop of the ring. While Eq. (2) describes the minimum mode volume of a conventional WGM or MRR for travelling wave resonance, it does not describe the true diffraction limited mode volume $V_{min}$ for a quasi-normal WGM. Any clockwise or counterclockwise travelling wave WGM can be described by the temporally delayed superposition of a single pair of degenerate symmetric and anti-symmetric quasi-normal modes. Considering the mode volume of the individual quasi-normal modes and maintaining the approximation of a homogenous medium under the approximation $n_{eff} \approx n$, the minimum diffraction limited mode volume of a standing wave WGM is given by:

$$V_{min} = \frac{1}{2}V_0 = \pi R \left(\frac{\lambda_0}{2n}\right)^2 = 2\pi m \left(\frac{\lambda_0}{2n}\right)^3 \quad (3)$$

Eq. (3) implies that standing wave resonances can exhibit a factor of 2 smaller mode volume than their travelling wave counterparts. In other words, MRRs based on standing wave modes inherently features a 2x enhancement in peak energy density, $U_e$, compared to travelling wave modes, coinciding with the formation of localized peaks and nulls in the time averaged field intensity as shown in Fig. 1C. Along this vein, breaking degeneracy and exciting individual standing wave modes has previously been pursued to increase sensitivity of WGM nanosensors[31,32] and the energy efficiency of MRR modulators[23].

Eq. (3) describes the minimum diffraction limited mode volume for a WGM resonator, which is fundamentally limited by the mode order, wavelength, and the resonator refractive index. A detailed mapping of these parameters and their influence on $V$ is illustrated in Fig. 1D. Notably, Eq. (3) indicates that the minimum diffraction limited mode volume scales in proportion to $mn^{-3}$. Eq. (3)'s linear scaling with mode order $m$ implies that high $Q$ factor resonators, which are often designed to operate with large bending radii and large $m$, typically exhibit large $V$. In practice, high $Q$ WGMs and MRRs also trade-off increased $V$ when using larger modal area waveguides and/or lower index platforms.

Breaking the diffraction limit in WGM microresonators has previously been demonstrated in travelling wave-based hybrid photonic-plasmonic MRRs[15]. To our knowledge such devices have not yet exploited standing waves, which could allow a factor of 2x further reduction in $V$ compared to travelling waves. However, plasmonic approaches suffer from limited CMOS compatibility and increased optical losses due to absorption, which can substantially degrade the intrinsic $Q$ factor.

In this work, we introduce an all-dielectric MRR with a sub-diffractive mode volume as illustrated in Fig. 1E. Our MRR is constructed from a high index contrast waveguide with a subwavelength grating (SWG) comprised of a low index diamond shape of width $w_l$ and height $w_h$ inside each unit cell. The diamonds from adjacent unit cells are non-overlapping, which produces a narrow high refractive index bridge of width $w_b = P - w_l$, where $P$ is the SWG period. For TE polarization, this design produces a strong localized enhancement in electric field intensity $|E|^2$ within the dielectric bridge separating each diamond (Fig. 1F), which enables mode volume reduction below the diffraction limit.

Our design is inspired in part by recent advancements in high-index contrast all-dielectric nanophotonics, which have shown how the peak electric field energy density $U_e$ in a high index material, can be enhanced by leveraging interfacial boundary conditions to Maxwell's equations [33–38]. Firstly, continuity of the normal component of electric displacement $D$ can be utilized to enhance the electric field $E$ by a factor $\frac{\epsilon_{hi}}{\epsilon_{lo}} = \left(\frac{n_{hi}}{n_{lo}}\right)^2$, where $\epsilon_{hi}$ and $\epsilon_{lo}$ are the relative permittivities, and $n_{hi}$ and $n_{lo}$ are the refractive indices of the high and low index materials respectively. This enhancement is traditionally only realizable in the low index medium, leading to the so called 'slot effect' [39–41]. However, by locally structuring the medium also in the orthogonal direction, continuity of the tangential component of $E$ can be harnessed to bring this enhancement into the high index medium. This allows for a theoretical enhancement in the max[$U_e$] term of Eq. (1) by a factor approaching $\left(\frac{n_{hi}}{n_{lo}}\right)^4$ [33,34].

We emphasize an important distinction regarding physically realizable vs. non-physically realizable electric field enhancements in all-dielectric systems. It is well known that the macroscopic form of Maxwell's equations theoretically results in electric field singularities at perfect corners. If such a system were physically realizable it would imply via Eq. (1) that the lower bound for a sub-diffractive mode volume $V'_{min}$ approaches zero, which is certainly not realistic. Recently a "singular dielectric nanocavity" was proposed to exploit such a singularity as a means to reduce the mode volume[42]. However, other works have already recognized that infinite singularities are physically unrealistic and endeavored to characterize only physically realizable field enhancements and physically accessible mode volumes[33,35,43]. Along this vein, here we adapt the $\left(\frac{n_{hi}}{n_{lo}}\right)^4$ scale factor obtained from Refs. [33,34] to represent an approximate and physically meaningful upper bound for the local $U_e$ enhancement. As such, we estimate the theoretical lower bound for the mode volume of a sub-diffractive all-dielectric WGM resonator to be non-zero and approximated by:

$$V'_{min} \approx \frac{\pi m}{4}\left(\frac{1}{n}\right)^7 \lambda_0^3 \quad (4)$$

which assumes $n_{lo} = 1$ and an index contrast $\Delta n = n - 1$. All practically realizable all-dielectric systems should exhibit a mode volume larger than this minimum bound. In practice, structuring the waveguide will naturally break the approximation $n_{eff} \approx n$, and secondly only some fraction of the $\left(\frac{n_{hi}}{n_{lo}}\right)^4$ intensity enhancement is likely to be realizable.

Eq. (4) provides important insight into the design and scaling principles of ultra-low mode volume all-dielectric resonators. It indicates that the minimum sub-diffractive mode volume for all-dielectric nanostructured optics $V'_{min}$ scales in proportion to $mn^{-7}$ (Fig. 1I), whereas $V_{min}$ scales with $mn^{-3}$ for conventional diffraction limited optics (Fig. 1D). Principally, this sub-diffractive scaling extends beyond WGMs and MRRs and indicates all dielectric resonators confined in three dimensions, e.g. photonic crystal resonators, Mie resonators, etc., have a minimum theoretical mode volume that can scale with $n^{-7}$ instead of $n^{-3}$. Notably, $V \propto n^{-7}$ scaling has recently been observed in the simulation of inverse designed 3D-sculpted sub-diffractive nanocavities [44]. These results further highlight the importance of developing novel materials with exceptionally high refractive indices [45].

We introduce a generalization of the mode volume expressions found in Eqs. (2-4) which enables applicability to both diffractive and sub-diffractive WGMs in travelling or standing wave configurations by introducing a dimensionless scale factor $\eta$:

$$V = \frac{1}{\eta} \pi R \left(\frac{\lambda_0}{2n}\right)^2 \quad (5)$$

where $\eta = 1$ corresponds to the classical diffraction limit for quasi-normal modes in a homogenous medium ($V = V_{min}$ from Eq. (3)), $\eta < 1$ to a device operating above this diffraction limit ($V > V_{min}$), and $\eta > 1$ to a device operating below the diffraction limit – i.e. in the "sub-diffractive" regime ($V'_{min} < V < V_{min}$). For example, a device with $\eta = 30$ would exhibit a mode volume 30x below the diffraction limit. In practice $\eta$ can be computed according to $\eta = V_{min}/V$, per the definitions in Eq.(1) and Eq.(3).

Through the appropriate nanostructuring of a high index contrast waveguide and the use of standing wave modes, as proposed here, we predict that $\eta$ values on the order of ~$10^1$-$10^2$ are feasible in silicon based MRRs depending upon the use of SiO$_2$ or air-cladding. This represents a substantial enhancement compared to $\eta \leq \frac{1}{2}$ for a conventional travelling wave WGM. Depending on the interplay between the mode order $m$, the number of grating periods $N = 2\pi R/P$, and the even vs. odd symmetry of the particular standing wave resonance, a rich variety of WGM resonances with distinctive local field enhancements could be realized (Fig. 1G,H).

## 3 Metawaveguide Simulation

Before investigating the resonant properties of the proposed metawaveguide MRR, we first simulate the properties of a straight metwaveguide, similar to the bus waveguide illustrated in Fig. 1E. Simulation of the metawaveguide is performed using the 3D finite difference time domain (FDTD) method (Tidy 3D) with results reported in Fig. 2. This simulation serves two purposes: (1) it provides a direct measurement of the electric field local enhancement relative to a conventional strip waveguide, allowing for estimation of $\eta$; and (2) it enables verification of the SWG waveguide properties and assessment of the efficiency of the strip-to-SWG mode convertor. Lastly, we note that the bus waveguide design illustrated in Fig. 1E will readily facilitate phase matching with the ring resonator waveguide, as in other types of SWG MRRs [46–50].

To mitigate losses when interfacing with the metamaterial waveguide, a strip-to-SWG waveguide mode convertor is introduced. The mode converter gradually tapers the diamond dimensions from $w_{l,min}$ x $w_{h,min}$ = 60 x 80 nm up to their nominal dimensions $w_l$ x $w_h$ = ($P$ - $w_b$) x $w_h$ nm, over a taper length $L_{taper}$. To preserve sub-wavelength operation and suppress taper reflections as the effective index varies with diamond size, the grating period inside the taper must also be chirped from $P_{min}$ up to $P$ over the same length. Here, we consider a fully etched 220 nm silicon device layer thickness and set the outer silicon waveguide width w = 800 nm, diamond height $w_h$ = 400 nm, SWG period width P = 305 nm, $P_{min}$ = 180 nm, $L_{taper}$ = 4.3 μm (20 unit cells), and examine a range of bridge widths $w_b$ from 4 nm to 30 nm. All diamond corners are rounded with a radius of curvature $r$ = 10 nm to eliminate corner singularities and provide physically realistic results [35].

Fig. 2A reveals the time averaged electric field distribution for 1550 nm light propagating through a SiO$_2$ cladded diamond metawaveguide with $w_b$ = 4 nm. Compared to the homogenous strip-waveguide, local field enhancements in silicon $|E|/|E_0| \approx 3.9$ are observed in the dielectric bridge between adjacent diamonds. This results in ~15x enhancement in peak electric field intensity $|E|^2$ and energy density $U_e$ in silicon compared to the unpatterned strip waveguide (Fig. 2B). As shown in Fig. 2C, at 1550 nm this device operates as a SWG and the strip-to-metawaveguide mode convertors exhibit high efficiency. Hence, the observed electric field intensity and energy density enhancement are non-resonant and broadband in nature. The enhancement increases to ~40x when the index contrast is increased by using an air cladding and is a strong function of the dielectric bridge width $w_b$ as illustrated in Figs. 2B,D. Visualization of the local field enhancement $|E|/|E_0|$ in the dielectric bridge for two bridge widths, $w_b$ = 4 nm and 30 nm, is shown in Fig. 2 E, F.

The simulation results in Fig. 2D provide a direct measure of the enhancement in peak electric field energy density relative to the strip waveguide. Since these enhancements stem from near-field engineering and not interference effects, they can be used to estimate the mode volume scaling factor $\eta$ from Eq. (6) when such waveguides are used to construct the MRR. At 1550 nm the reference strip waveguide exhibits a mode area $A_n = 0.08~\mu m^2 = 1.63 \left(\frac{\lambda_0}{2n}\right)^2$, resulting in $\eta = 0.307$ for a conventional travelling wave MRR. The metawaveguide MRR, meanwhile, signficantly enhances $\eta$ through a combination of near field enhancement (Fig. 2D) and the use of standing wave modes to yield mode volumes below the diffraction limit ($\eta > 1$). Our results indicate an oxide or air cladded metawaveguide MRR could achieve $\eta \approx 8$ or 24 respectively, which represents up to an ~80x reduction in mode volume compared to a conventional silicon MRR.

## 4 Metawaveguide Ring Resonator Simulation

To verify the large $\eta$ and sub-diffractive mode volume of the proposed metawaveguide MRR architecture, we performed 3D FDTD simulation. Specifically, we model an oxide cladded metawaveguide MRR excited from the bus waveguide as illustrated in Fig. 1E, using the same taper mode convertor design as in Fig. 2. Accurate determination of the mode volume, $V$, requires a fine mesh resolution to accurately capture the localized field enhancements. However, this presents a practical computational challenge for large area and high resolution 3D FDTD simulations, which can be prohibitively expensive in terms of compute and memory requirements. Here we leverage the computational power of the GPU-cluster environment utilized by Tidy3D (Flexcompute Inc.) to perform 3D FDTD in a simulation domain with x, y, z span of 45 μm × 68 μm × 2 μm. We employed symmetric boundary conditions in the z-axis, an automated conformal mesh with a maximum grid resolution of $\lambda_0/15$, a custom mesh override to a minimum resolution of 5 nm in a portion of the xy plane, and sub-pixel averaging resulting in a simulation requiring 122M grid points. In the simulation results reported in Fig. 3, the key geometrical parameters were: $R = 19.4175$ μm, $N = 2\pi R/P = 400$, $P \approx 305$ nm, $w_b = 10$ nm, and a coupling gap, $g = 175$ nm.

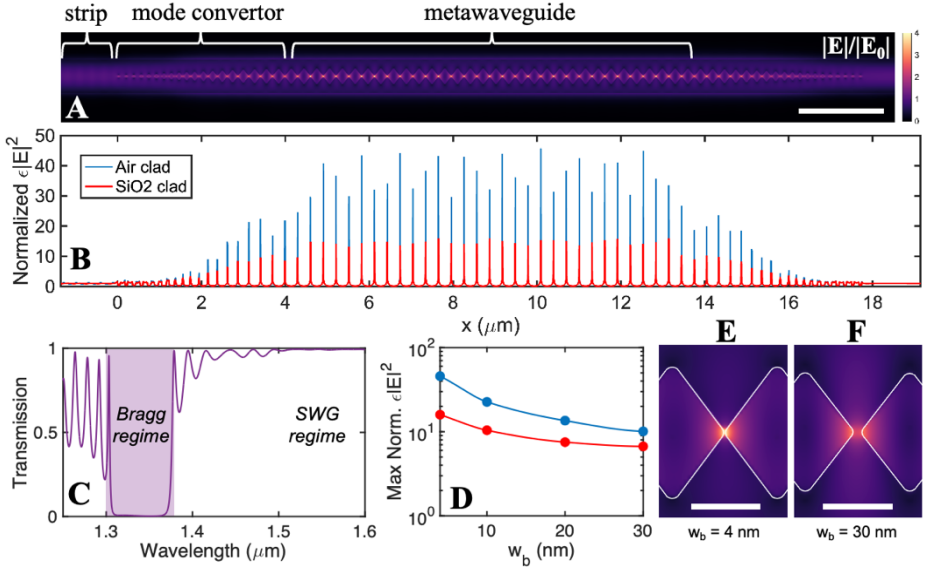

**Figure 2: Optical properties of the subwavelength diamond metawaveguide.** (A) Time averaged electric field distribution of 1550 nm light propagating through a $SiO_2$ cladded diamond metawaveguide (including input/output mode convertors). The field is normalized to the amplitude in the feeding strip waveguide. (B) Corresponding electric field energy density enhancement along the center of the metawaveguide for $SiO_2$ and air cladded devices with $w_b$ = 4 nm. (C) Transmission spectrum of the device considered in (A), revealing high transmission efficiency and subwavelength operation at 1550 nm. (D) Summary of the peak electric field energy density enhancement as a function of bridge width and cladding (blue = air, red = $SiO_2$). (E) Zoomed view of the time averaged electric field distribution in a single unit cell from part (A) with $w_b$ = 4 nm, and comparison to (F) a device with $w_b$ = 30 nm [same color scale as in part (A)].

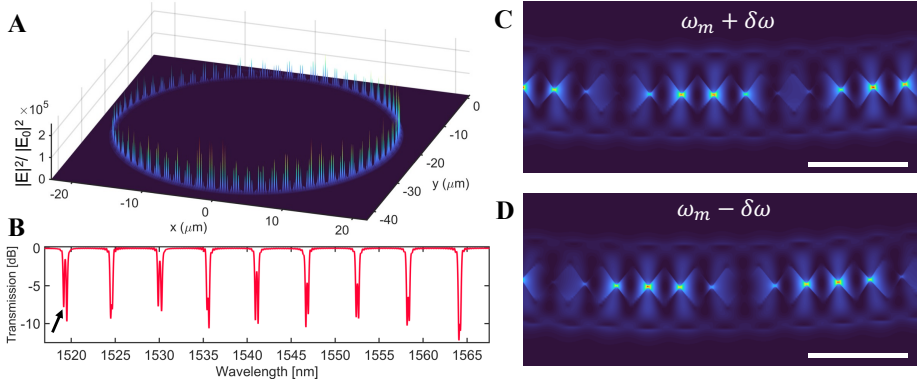

**Figure 3:** (a) 3D FDTD simulation of a metawaveguide MRR cavity with R = 19.4 μm showing the resonant electric field intensity at $\lambda_0 = 1519.14$ nm. (b) Simulated transmission spectrum through the bus waveguide, revealing mode splitting. (c,d) Zoomed near field intensity visualization for the two standing wave resonances associated with mode splitting at $\lambda_0 = 1519.14$ nm and 1519.5 nm (scale bar = 1 μm).

Fig. 3A shows the simulated electric field intensity distribution for a resonant mode at 1519.14 nm. The near field corresponds to a standing wave resonance with characteristic peaks and nulls. The mode volume is computed to be $V = 0.128\ \lambda_0^3$, which corresponds to $\eta = V_{min}/V = 6.41$. This confirms the deeply sub-diffractive nature of the mode volume and is in excellent agreement with the predicted $\eta = 6.39$ based on results in Fig. 2D. The resonant mode order is estimated to be near $m \approx 160$, indicating a modal effective index near $n_{eff} \approx 2$. This results in $M \approx 320$ intensity peaks which are superimposed on the $N = 400$ unit cells. The periodic dielectric function of the MRR lifts the degeneracy of the symmetric and anti-symmetric modes, resulting in mode-splitting which is visible in the simulated transmission spectrum (Fig. 3B). Corresponding near field visualizations shown in Fig. 3C, D show that each mode exhibits a unique mode pattern. Interplay between the mode pattern and the periodic nanostructures gives rise to a beat pattern in the sub-diffractive electric field enhancement. Hence, the maximum $|E|^2$ or $U_e$ occurs periodically with a beat frequency of $|N - M| \approx 80$ occurrences per circumference. We note that $M < N$ is necessary to remain in the SWG regime and away from the photonic bandgap. In this regime, selecting the resonant mode order ($m = M/2 = 2\pi R n_{eff}/\lambda_0$) provides freedom to manipulate the beat frequency and the distribution of hot spots, while choosing between the $\omega \pm \delta\omega$ modes can provide additional freedom to spatially shift or modulate the mode pattern as in Fig. 3 C,D. When $M = N$ the device enters the Bragg regime. If desired, it would be possible to harness a dielectric band edge resonance to achieve maximum $|E|^2$ within every unit cell [51].

## 5 Experiment & Discussion

Scanning electron microscope images of fabricated diamond metawaveguide MRRs are shown in Fig. 4. These devices were fabricated in a standard 220 nm silicon on insulator (SOI) platform using the NanoSOI MPW fabrication process by Applied Nanotools, Inc. and subsequently measured at The University of British Columbia through their Silicon Electronic-Photonic Integrated Circuits program [52]. SEM images were captured prior to embedding the devices in a ~2.2 μm thick top $SiO_2$ cladding by plasma enhanced chemical vapor deposition. As shown in Fig. 4A and 4B, we designed MRRs with two different bridge widths, $w_b$, which were estimated by SEM image analysis to be ~14 nm for the narrower design and ~25 nm for the wider design respectively. The SWG period was set to 305.4 nm and implemented by using an integer number of unit cells, $N = 720$, in MRRs with radius $R = 35$ μm. To aid experimental study of the metawaveguide MRR's intrinsic resonant properties and Q factor, devices were replicated with four different gaps, $g = \{175, 200, 250, 300\}$ nm.

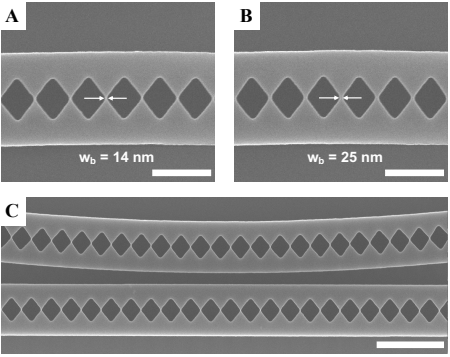

**Figure 4: SEM images of the fabricated devices.** (A) Microring resonator with narrow bridge width $w_b = 14$ nm and (B) wider bridge width $w_b = 25$ nm [scale bars = 500 nm]. (C) Bus waveguide coupling section with a 175 nm gap [scale bar = 1 μm].

An experimental transmission spectrum for a selected device with $R = 35$ μm, $w_b = 25$ nm, and $g = 300$ nm is shown in Fig. 5. The results reveal high Q factor resonances with an average loaded $Q = 2.71 \times 10^4$. The average free-spectral range was measured to be 2.75 nm, corresponding to a metawaveguide group index $n_g \approx 3.83$ near 1550 nm. Resonant mode splitting is also observed, with an average mode split of 0.175 nm. Whereas mode splitting in MRRs is often interpretted as a sign of fabrication defects, here it is an expected and important feature for minimizing the resonator mode volume.

To achieve a pure standing wave excitation, the mode splitting magnitude should be significantly greater than the linewidth. If the mode splitting is comparable to or less than the linewidth, the resonator excites partial standing waves or travelling waves. To gain insights into the modal properties of our metawaveguide MRRs, we analyzed resonator mode splitting as a function of linewidth for each waveguide gap: $g = \{175, 200, 250, 300\}$ nm, as shown in Fig. 5C-F. This analysis allows for direct visualization of the prevalence of standing, partial standing, and travelling wave WGMs. Correlation analysis between mode splitting and linewidth can also be used as a tool to probe for a mechanistic link between the two parameters. In the case of ultra-high Q resonators where defects simultaneously increase losses and break structural symmetries a strong correlation between mode splitting and linewidth has been observed [12]. In our case however, the structural symmetry is broken by the SWG and the linewidth is limited by other factors affecting radiation losses (to be discussed). As shown in Fig. 5C-F, a weak correlation between mode split and linewidth, $C = 0.381$, is observed for $g = 175$ nm. As the coupling gap is increased to $g = 300$ nm, the average linewidth decreases and the correlation vanishes to a negligible value, C = -0.008. This confirms the mode splitting and linewidth (Q factor) are independent, especially for $g \geq 250$ nm. In the case of small coupling gaps ($g < 250$ nm) the modest correlation, C = 0.266 to 0.386, is likely attributable to increased interaction between the ring resonator and the bus waveguide. For $g = 300$ nm the prevalence of standing wave modes is high with ~77% of resonances exhibiting mode splitting greater than 1x the linewidth and ~60% greater than 2x the linewidth. This confirms that a majority of the resonant modes can be classified as standing wave modes which exhibit high $\eta \approx 6$.

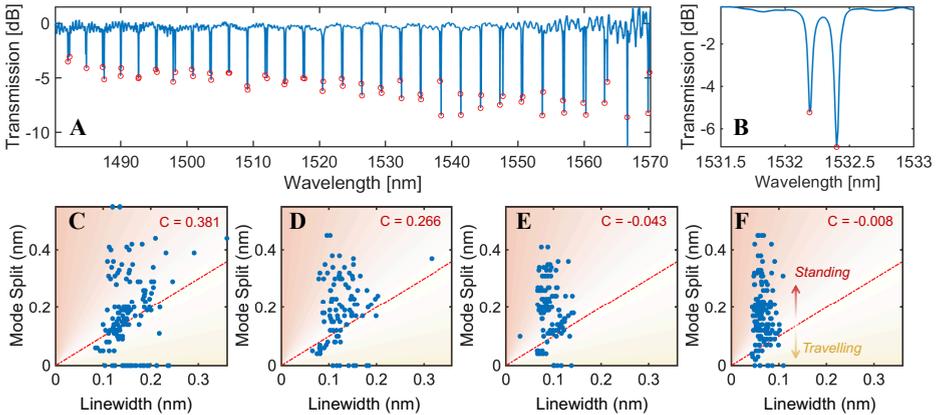

**Figure 5: Spectral characterization of a metawaveguide MRR.** (A) Experimentally measured transmission spectrum for a device with $R = 35$ μm, $g = 300$ nm, and $w_b = 25$ nm. (B) Zoomed view of the spectrum near 1532 nm, revealing mode splitting on the order of 0.2 nm. (C) Resonant mode splitting vs. linewidth inclusive of both $w_b = 14$ nm and 25 nm for $g = 175$ nm, (D) $g = 200$ nm, (E) $g = 250$ nm, and (F) $g = 300$ nm.

Our simulation (Fig. 3) and experimental results (Fig. 5) confirm that all-dielectric MRRs can operate with both low $V$ and high $Q$. A natural next step is to determine the key factors limiting both $V$ and $Q$ and to determine if there is a path to improving them and scaling Q/V in future work.

With respect to $V$ reduction, we reiterate that the most crucial degrees of freedom are in the refractive index contrast, nanostructure geometry, and usage of standing wave vs. travelling wave modes.

With respect to quality factor, our analysis indicates that the total loaded $Q$ is presently primarily limited by: coupling losses and radiative bending losses, while scattering based propagation losses are not yet a limiting factor. In Fig. 6 we fit the experimentally measured (loaded) $Q$ factor data to a loaded ring resonator model to extract estimates for the maximum and average intrinsic quality factor, $Q_i$[53]. For devices with $w_b = 25$ nm we obtain $Q_{i,avg} = 4.11 \times 10^4$ and $Q_{i,max} = 8.23 \times 10^4$ which are limited by bending dominated resonator losses of 0.056 dB/rad (16 dB/cm) and 0.028 dB/rad (8 dB/cm) respectively.

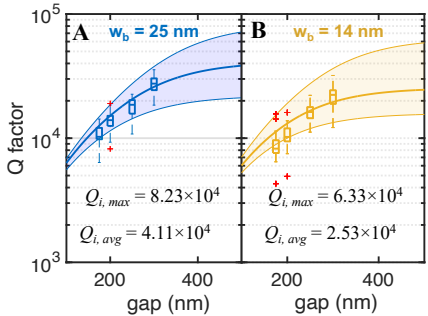

**Figure 6: Q-factor analysis.** (A) Experimental and fitted trend in loaded Q-factor vs. coupling gap between the MRR and bus waveguide for $w_b = 25$ nm, and (B) $w_b = 14$ nm devices with ring radius 35um. This analysis enables extraction of the average and maximum intrinsic quality factors (as indicated).

For devices with $w_b = 14$ nm the larger diamond size lowers the waveguide effective index and group index slightly ($n_g \approx 3.73$ vs. 3.83), coinciding with elevated bending losses estimated to be 0.0875 dB/rad (25 dB/cm) and 0.035 dB/rad (10 dB/cm) for $Q_{i,avg} = 2.53 \times 10^4$ and $Q_{i,max} = 6.33 \times 10^4$ respectively.

In our present design, the diamond metawaveguide MRR achieves localized field enhancements at the cost of substantially lower effective index ($n_{eff} \approx 2$) vs. a conventional strip waveguide ($n_{eff} \approx 3$). In the future, we anticipate the SWG unit cell design can be further optimized to maintain strong field enhancements approaching the $\left(\frac{n_{hi}}{n_{lo}}\right)^4$ approximate upper bound while preserving higher $n_{eff}$, lower bending losses, and higher Qs. These results also highlight an important point which is often overlooked when discussing mode volume $V$. Per Eq. (1), the mode volume provides an inverse measure of the peak electric field energy density max[$U_e$] normalized to the total energy confined in the resonator $\int U_e dV$. The sub-diffractive phenomena at play here, primarily functions by locally enhancing max[$U_e$]. As such it would be potentially misleading to suggest all-dielectric sub-diffractive mode volumes necessarily exhibit improved "confinement", when in fact in many cases the majority of the resonator energy is not more tightly confined. Instead it would be more precise to describe sub-diffractive mode volumes as exhibiting locally improved optical "concentration" [54]. Clarity in this interpretation is important, since the applications that would benefit most strongly from the sub-diffractive mode volumes discussed here, and in other related works, are precisely those applications that demand or benefit from *locally* improved optical concentration. Leading examples of such applications include enhancing light-matter interactions with nanoparticles [31], single molecules [55], atoms [19], ions [56], and individual quantum emitters [20–22,57].

## 6 Conclusions

In this paper, we have reported an approach to achieve deeply sub-diffractive mode volumes in nanoengineered all-dielectric ring resonators. Our approach leverages simultaneous usage of subwavelength metamaterial waveguides, also known as SWGs[58] or metacrystals[59], along with the excitation of standing wave rather than travelling wave resonances. This principally enables the mode volume to be reduced by a factor of $2\eta$ compared to a diffraction limited travelling wave WGM. In our present system comprised of $Si/SiO_2$ we achieve $2\eta \approx 12$, representing >10x reduction in modal volume. Notably, the combination of diffractive mode volume scaling, $\sim m\lambda_0^3 n^{-3}$, and sub-diffractive intensity enhancement scaling, $\sim \left(\frac{n_{hi}}{n_{lo}}\right)^4$, indicates a lower limit for sub-diffractive all-dielectric mode volumes which scales in proportion to $m\lambda_0^3 n^{-7}$. This WGM architecture supports many design degrees of freedom which can be used to further tailor or optimize $V$ and/or $Q/V$ in future work. These features can enable enhanced light-matter interactions in the near-field of WGM and resonators and MRRs may prove especially advantageous for use in emerging quantum photonic, nano-sensing, and nano-manipulation architectures.

**Authors' statements**

**Acknowledgments:** We thank Clemson University for their generous compute time allocation on the Palmetto cluster. We acknowledge L. Chrostowski, I. Taghavi, and O. Esmaeeli and the edX UBCx Phot1x Silicon Photonics Design, Fabrication and Data Analysis course, which is supported by the Natural Sciences and Engineering Research Council of Canada (NSERC) Silicon Electronic-Photonic Integrated Circuits (SiEPIC) Program. Devices were fabricated through the NanoSOI MPW fabrication process by Applied Nanotools Inc. **Research funding:** Air Force Office of Scientific Research (AFOSR) Young Investigator Research Program (G. Pomrenke), FA9550-19-1-0057. National Science Foundation (NSF) Award #2235443.

**Author contribution:** All authors have accepted responsibility for the entire content of this manuscript and approved its submission.

**Conflict of interest:** J.R. acknowledges U.S. Patent US11320584B2.

**Data availability statement**: Data supporting this research are available upon reasonable request.